\newcommand{\rom}[1]{\uppercase\expandafter{\romannumeral #1\relax}}
\documentclass[aps,pra,twocolumn,superscriptaddress]{revtex4-1}
\usepackage{graphicx}
\usepackage[colorlinks=true,citecolor=blue, urlcolor=blue, pdfborder={0 0 0}]{hyperref}
\usepackage{amsmath}
\usepackage{amssymb}
\usepackage{bbold}
\usepackage{subfigure}
\usepackage{epsfig} 
\usepackage{epstopdf}
\usepackage{float} 
\usepackage{tikz}
\usepackage[percent]{overpic}
\usepackage{comment}
\usepackage{placeins}
\usepackage{multirow}
\usepackage{rotating}
\usepackage{hhline}
\usepackage{xcolor}

\begin{document}
\title{Measurement-based cooling of many-body quantum systems}

\author{Tarek A. Elsayed}
\email{tarek.ahmed.elsayed@gmail.com}
\affiliation{Department of Physics, School of Science and Engineering, The American University in Cairo, AUC Avenue, P.O. Box 74, New Cairo, 11835, Egypt}

\date{\today}

\begin{abstract}
We introduce a novel technique for efficiently cooling many-body quantum systems with unknown Hamiltonians down to their ground states with a high fidelity. The technique involves initially applying a strong external field followed by a sequence of single-degree-of-freedom (single-qubit) measurements and radiofrequency (RF) pulses to polarize the system along the field  direction. Subsequently, the field is adiabatically switched off, allowing the system to evolve towards its ground state as governed by the quantum adiabatic theorem. 
 We present numerical simulation results demonstrating the effectiveness of the technique applied to quantum spin chains with long-range and short-range interactions as prototypes for many-body quantum systems.

\end{abstract}

\maketitle

\section{Introduction}

The problem of driving a quantum system from a typical initial state to a desired final state is of enormous practical importance and falls within the subject of quantum control. With the advent of quantum technologies that  rely heavily on quantum systems being in certain states (e.g., the ground state), the field of quantum control has witnessed a surge of interest in the past two decades \cite{lloyd2000,reiserer2013,magrini2021}. The essence of optimal control or quantum state engineering \cite{schirmer2007,chiu2018,makhlin2001} is to design the (time-dependent) Hamiltonian that coherently drives the quantum system from an initial state to a target state with  high fidelity (open-loop control). Another important paradigm of quantum control that derives from classical control theory \cite{sze} is to use a feedback control loop that involves performing a measurement of the quantum system and using the measurement outcome to control the parameters of that quantum system (closed-loop control). Both types of control find important applications in many areas \cite{koch2022} such as quantum computing \cite{verstraete2009}, quantum sensing \cite {titum2021} and quantum communications \cite{pastorello2016}.

Controlling the dynamics of a quantum system with a feedback mechanism typically involves monitoring a few parameters of the system and uses the feedback loop in order to stabilize the state of the system (e.g., against the effects of the environment) \cite{dong2010}. A major problem in quantum control is the unavoidable back-action entailed by the measurement on the system. This problem could be mitigated by evading the effect of the back-action \cite{vanner2013,rossi2018}. Recently, however, other schemes have been proposed that actually harness the effect of the projective measurement on the system and make use of the back-action as an ingredient of the algorithm itself \cite{rossini2020,roy2020,dasari2021,puebla2020,sorensen2018,mazzucchi2016,pedersen2014,ashhab,buffoni2019}. 
The recent progress in quantum measurement techniques \cite{zhang2017}, and the development of non-invasive measurement techniques that inflict minimal perturbation to the measured system such as weak measurement \cite{white10} or  local measurement of individual particles in a many-particle system \cite{gross2021} open new possibilities in utilising the feedback control loop approach in quantum state engineering. The purpose of this work is to present a new technique in this regard that relies on the measurement of single degrees of freedom (i.e., single qubits involving individual particles or small groups of particles) and employing the measurement outcome in a feedback loop in order to steer a many-body quantum system from a random initial state (typically a high temperature state) toward its ground state. We use quantum spin chains with short-range and long-range interactions as prototypes for many-body quantum systems.

 An efficient method that is often used for cooling down a quantum system close to its ground state is adiabatic demagnetization \cite{mirasola2018} which requires a highly polarized initial state. However, bringing the system to this state in the first place is a challenge in itself, especially if the system is in contact with an environment. A system in thermal equilibrium in a strong magnetic field will typically have  a very small polarization determined by the temperature and the magnetic field strength. The algorithm we propose gradually polarizes the system by means of a feedback loop that involves applying a sequence of single-qubit measurements and radio frequency (RF) perturbations to the system. It makes use of the fact that it is much easier to polarize small parts of the system, one at a time, thus imparting a minimal disturbance to the rest of the system, instead of polarizing the whole system in one shot. The reason is that the size of the subset of the Hilbert space corresponding to one partition of the system being completely polarized is much larger than the size of the subset corresponding to the whole system being completely polarized and therefore it is easier to drive the system into the states corresponding to polarized individual partitions, one partition at a time.

\section{Measurement-based Cooling}
 In the proposed technique, we adiabatically evolve a  quantum system of interacting many particles towards its ground state after polarizing it by applying a sequence of measurements and RF perturbations. Therefore, the algorithm consists of two steps:
 
  In Step \rom{1}, an external strong field $\boldsymbol{B}$  which alters the energy spectra and eigenstates of the system is applied, letting the ground state of the new Hamiltonian have a very large polarization anti-parallel to $\boldsymbol{B}$. We then make a sequence of spin polarization measurements of randomly chosen individual particles or subsets of particles along the direction of the external field. If the magnetization of the measured partition is aligned anti-parallel to $\boldsymbol{B}$, we do nothing. Otherwise, if it is aligned parallel to $\boldsymbol{B}$,  we perturb the system by an RF field perpendicular to $\boldsymbol{B}$ for a certain time interval before we make another measurement of the same partition. We keep measuring the same partition repetitively at specific times separated by periods of unitary evolution under the effect of the RF perturbation till it has been projected onto the correct direction. If we repeat this procedure many times (much more than the total number of partitions) before the system relaxes back to thermal equilibrium, the system will eventually be polarized along  $\boldsymbol{B}$, and thus its entropy will be hugely reduced. Of course, while measuring a certain particle and perturbing the system, the state of a previously measured particle will be altered. Nevertheless, our numerical simulations show that, by applying this scheme, a small-sized quantum system will always find a path towards the fully polarized state. Once fully polarized, the system will be very close to the ground state of the new Hamiltonian.

In Step II, we do an adiabatic depolarization, which is a well-known technique used for attaining very low temperatures by switching off the external field very slowly \cite{giauque1933,de1933,shepherd1965}. According to the quantum adiabatic theorem \cite{bachmann2017} and under certain conditions about the system, the state of the slowly varying system will remain in the vicinity of the ground state of the instantaneous Hamiltonian. Eventually, when the field is completely turned off, the system ends up very close to its true ground state. Note that applying this scheme does not require knowing the exact Hamiltonian of the bare system. 

The most prominent application of quantum control is the optimal control of solid state Nuclear Magnetic Resonance (NMR) which has been actively pursued for a long time \cite{chuang2005,khaneja2005,nielsen2007,ram2022}. In the following analysis, we shall assume an NMR setting of small quantum spin chains as a prototype system, without loss of generality to other strongly-interacting many body quantum systems. We first employ projective measurements of single spins using single-particle detectors in  A, then we utilize bulk detectors which measure the direction of the magnetization of multi-particle subsets in B.

\subsection{Projective measurement of single particles}
Let us illustrate how this algorithm works using single-particle detectors through a detailed example. Consider an isolated quantum spin-1/2 chain consisting of $N$ particles with periodic boundary conditions and local nearest-neighbor interactions. The Hamiltonian for this system is given by:  \[\mathcal{H}_0=\sum_{m} S_m^x S_{m+1}^x-0.5(S_m^y S_{m+1}^y-S_m^z S_{m+1}^z)+0.3\sum_{m}S_m^y,\] where $S_m^i$ represents the spin operator for the m\textsuperscript{th} spin in the i\textsuperscript{th} direction (i = x, y, z). This system has a ground state energy  $E_0=-4.189  \text{ J}$  (we take $\hbar=1$ and $\gamma=1$ throughout this article, where $\gamma$ is the gyromagnetic ratio). The last term in the Hamiltonian is added to break the degeneracy of the ground state. The initial state of the chain is assumed to be an infinite-temperature pure state. A time step of $dt=0.001  \text{ s}$ is used in the simulation while a fourth-order Runge-Kutta method is used to evolve the time-dependent Schrodinger's equation \cite{elsayed2013}. 

 In Step I, we switch on an external magnetic field $\boldsymbol{B}$ in the $z$-direction of strength $B_0=10\  \text{T}$ which adds a Zeeman term $\mathcal{H}_z=B_0\sum_{m}S_m^z$ to the Hamiltonian. Our goal is to polarize the system anti-parallel to $\boldsymbol{B}$ and thus prepare it for the adiabatic demagnetization phase in Step II. Let us assume that we make a measurement of a particular spin along the $z$-axis and find it in the wrong direction, parallel to $\boldsymbol{B}$. In a typical NMR setting, the ideal method to perturb the system in order to flip the direction of a spin to a direction anti-parallel to $\boldsymbol{B}$ is to apply a targeted $\pi$-pulse to that particular spin. However, selective excitation of individual particles might not be practical for relatively large systems. As an alternative, we apply a weaker pulse to the entire system which does not perturb it much. In doing so, the repetitive measurement of a certain spin several times while applying the weak RF perturbation between the measurements will increase the likelihood to flip that spin while, at the same time, not perturb the other particles substantially. Our simulation shows that this technique works so well for systems of interacting quantum spins. 

We, therefore, employ a uniform RF field in the $x$-direction which affects all spins equally. This field adds a time-dependent term $h_x=h(t)\sum_{m}S_m^x$ to the Hamiltonian. The time varying function $h(t)$ is defined as $h(t)=h_0g(t)\cos(\omega t)$ where $\omega=5$ rad/s (i.e., half the Larmor frequency defined as $\gamma B_0$) represents the frequency of the RF field and $h_0$ represents the perturbation strength (set to 1 in our case). The function $g(t)$ is an on-off switch equal to 1 or 0 depending on whether we switch on the RF field or not. We perform a measurement of a random spin along the $z$-direction every period $T$.  If the spin is found to be pointing down  (desired state), we select another randomly chosen spin in the next round. Otherwise, if we find the spin pointing up along $\boldsymbol{B}$  (undesirable state), we switch on the RF perturbation and keep measuring the same spin  every $T$. We take $T$ to be equivalent to 0.5 cycles of the RF field. We continue this process until a large number of consecutive measurements consistently show all measured spins to be in the down state. This indicates a high probability of the system achieving full anti-parallel polarization (a measurement of the total magnetization can be done in this case for confirmation). 


Each measurement in our scheme projects the state of the system onto a subspace of the full Hilbert space of size $2^{N-1}$. Ideally, measuring one single degree of freedom should introduce only minimal disturbance to the system overall. 
While it is intuitive that the average time for the downfall to the fully polarized state increases as the size of the Hilbert space increases, the precise scaling of this average time with the number of spins remains an open question.
Since the proposed scheme involves measuring one particle at a time,  the time taken to polarize the majority of spins will at best scale linearly with the size of the system. It should be noted that the proposed scheme is model-independent, i.e., it does not require prior assumptions about the system or an initial learning phase that scales exponentially, as seen in some existing methods  \cite{negrevergne2006,bukov2018}.

For this algorithm to function correctly, the average time taken to flip a spin that was measured to be in the wrong direction should be sufficiently short such that that the perturbation caused by the RF pulses does not increase the probability to flip another spin which was already measured in the desired direction beyond 50\% when it is re-measured. This requirement imposes a further constraint on the maximum system size that can be effectively cooled using this method since the larger the system, the more perturbations  a correctly polarized spin will encounter before it is re-measured. Additionally, for a system in contact with a thermal reservoir, the time taken to fully polarize the system should be shorter than the relaxation time to equilibrium from the completely polarized state, which is typically of the order of the $T_1$ relaxation time constant.

\begin{figure}[t!] \setlength{\unitlength}{0.1cm}
\begin{picture}(60 , 90 )
{
\put(-10, 45){ \includegraphics[scale=0.25]{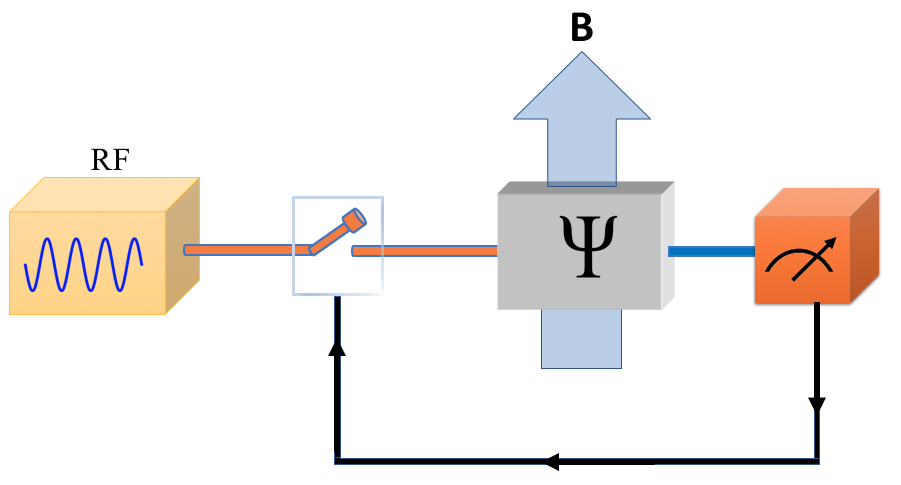}}
\put(-0, 0){\includegraphics[scale=0.5]{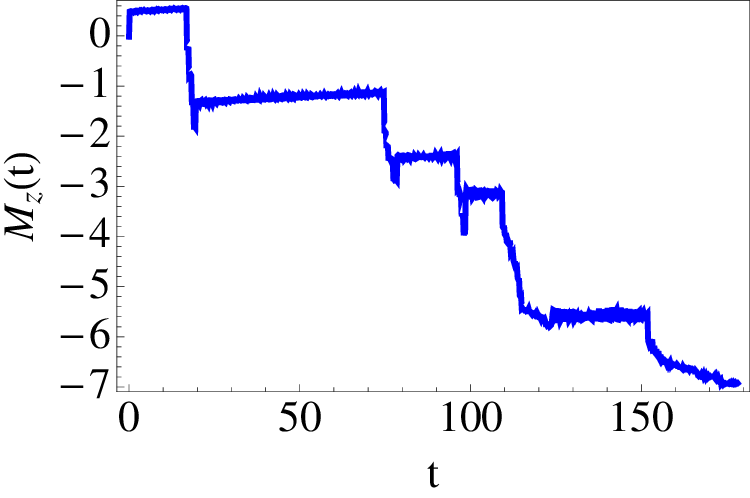}}

\put(-10, 85){\text{(a)}}
\put(-10, 45){\text{(b)}}
}
\end{picture}
\caption{ (a)  {\bf Measurement-Based Quantum Control Circuit:} 
Schematic of the proposed quantum control circuit. 
The  measurement outcomes from an apparatus measuring single degrees of freedom are used to control the perturbation of the quantum system by RF pulses. A strong external field $\boldsymbol{B}$ is switched on during this process. (b) {\bf Magnetization Dynamics During Polarization: } The evolution of the total magnetization of a system consisting of 14 spin-1/2 while applying the measurement-based polarization algorithm, starting from an unpolarized state.}
\label{fig-concept}
\end{figure}

In Fig. 1, we show the  $z$-component of the total magnetization $M_z\equiv \langle \hat{M}_z \rangle$, where $\hat{M}_z=\sum_{m} S_m^z$, during the downfall to the fully polarized state for a system consisting of $N=14$ spins. In general, $M_z(t)$ will keep fluctuating randomly due to the back-action entailed by the measurement and the continuous perturbation of the RF field before it eventually embarks on a  ``free-fall'' trajectory to the desired fully polarized state under the successive acts of projective measurement. While the exact time taken for this event to take place is unpredictable, the spontaneous downfall of the state of a small quantum system to the desired state seems to be inevitable.

The essence of the proposed algorithm is to create a completely stable state (the desired fully polarized state) and a completely unstable state (the state fully polarized in the opposite direction). The more the system approaches the undesired state, the more corrective RF pulses will be applied to it that drives the system away from that state. Our simulations show that, sooner or later, and no matter how large the size of the Hilbert space is, the system will eventually be ``attracted'' to a free-falling path that drives it steadily towards the desired state by this corrective mechanism. The stronger the external field with respect to the typical interaction constants of the bare Hamiltonian $\mathcal{H}_0$, the more stable the desired state will be since $\hat{M}_z$ commutes with $\mathcal{H}_z$ but not with $\mathcal{H}_0$. 
The high polarization obtained in Step I of the proposed technique can be combined with usual NMR techniques to achieve a much better resolution in NMR spectroscopy of small spin clusters.

As mentioned earlier,  the fully polarized state in the presence of a strong external field is very close to the ground state of the full Hamiltonian $\mathcal{H}_0+\mathcal{H}_z$. In our case, the ground state energy for $\mathcal{H}_0+\mathcal{H}_z$ is  \mbox{-68.39 J} while the energy of the fully polarized state is \mbox{-68.25 J}.  In Step II of the technique, we switch off the field very slowly by reducing its strength according to an exponential decay function: $B(t)=B_0*\exp(-t/T_0)$ with $T_0=10^4  \text{ s}$ . In Fig. 2, we show the evolution of the energy of the bare system as $\boldsymbol{B}$ is gradually switched off and the fidelity of the instantaneous state $|\psi(t)\rangle $ with respect to the exact ground state $|\psi_0\rangle$, which is  defined as $|\langle \psi_0 | \psi(t) \rangle|^2$. The achieved energy at the end of the adiabatic evolution approaches \mbox{-4.152 J}, which differs from the exact ground state energy by less than 1\% while the fidelity approaches 93\%. In general, the time required by the adiabatic evolution phase  depends on the energy gap of the system. It has been shown that this time grows polynomially with the size of the system when the system exhibits certain symmetries \cite{farhi2000}.

\subsubsection{Control Parameters}

There are a few control parameters that can be adjusted in this algorithm for achieving the best performance. The strength of the external field is taken to be one order of magnitude stronger than the typical interaction strength of the bare Hamiltonian. Increasing the field strength further can enhance fidelity, but, on the other hand, will prolong the adiabatic evolution time in Step II. The strength of the RF pulses is taken to be of the same order of magnitude as the interaction strength between the particles. Using much stronger RF pulses will have the side effect of strongly disturbing the spins which have already been polarized in the desired direction, whereas much weaker pulses will not cause enough disturbance to the spin that has been measured to be aligned in the wrong direction.

In typical NMR experiments, the strength of the external field is much stronger than the  local field at each particle resulting from its interaction with its neighbors. In this case, the appropriate RF frequency to be used to perturb the system is the Larmor frequency. In our numerical simulation, where the external field is only an order of magnitude stronger than the local field,  we take the RF frequency to be a few multiples of the typical interaction constant of $\mathcal{H}_0$.
We also set the time interval between two successive measurements of the same spin to be half a cycle of the RF field. This choice  ensures that the spin rotates in one single direction around the $x$-axis, thus maximizing the probability of flipping it during the next measurement event. However,  on the practical side, this choice presents a challenge concerning the required switching frequency. The probability to flip the spin in the $z$-direction  will increase when enough  polarization in the $xy$-plane of that particular spin  builds up. Allowing the rate of projective measurements to become too fast, compared to the timescales of the intrinsic dynamics, will not leave  sufficient room for spin polarization in the $xy$-plane to develop and, at the same time, will also let the quantum Zeno effect set in, thus freezing out the spin dynamical evolution. On the other hand, setting the measurement frequency  too low while applying the RF perturbation may substantially disturb the rest of the system by the RF field, and hence increase the probability of losing the polarization achieved so far.

\begin{figure}[t!] \setlength{\unitlength}{0.1cm}
\begin{picture}(60 , 90 )
{
\put(-5, 45){ \includegraphics[scale=0.5]{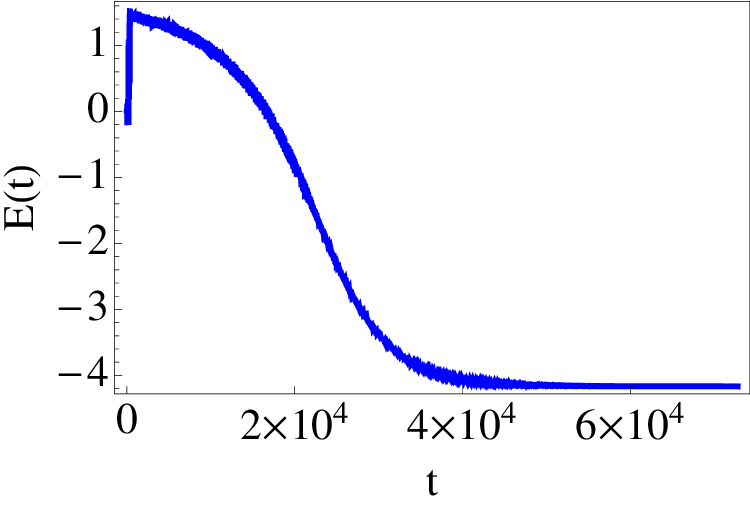}}
\put(-5, 0){\includegraphics[scale=0.5]{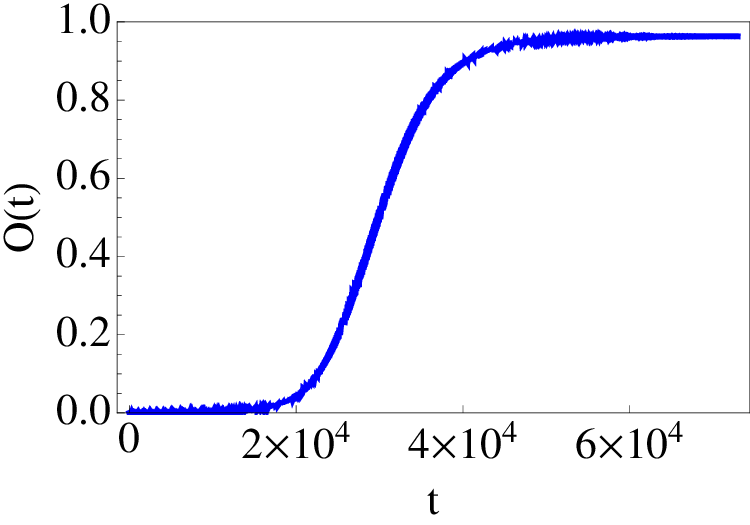}}

\put(25, 59){ \includegraphics[scale=0.25]{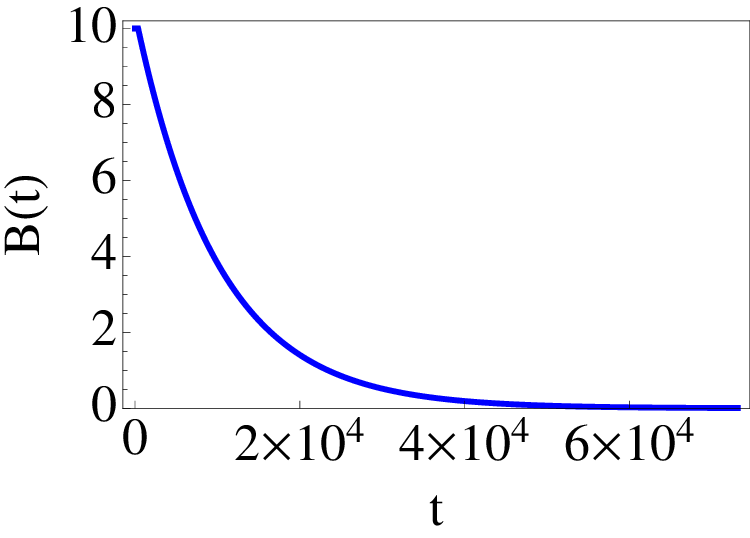}}
\put(27, 10){\includegraphics[scale=0.24]{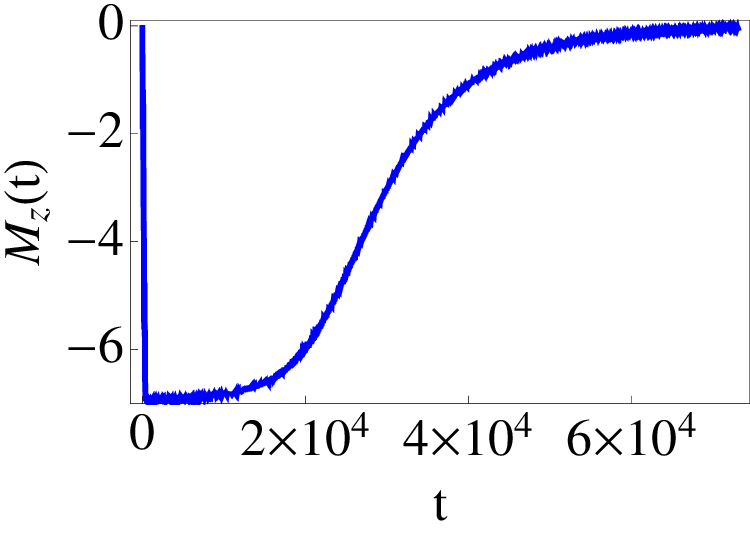}}

\put(25, 59){ \includegraphics[scale=0.25]{H.eps}}
\put(27, 10){\includegraphics[scale=0.24]{Z.eps}}

\put(-10, 85){\text{(a)}}
\put(-10, 45){\text{(b)}}

\put(40, 83){\text{(c)}}
\put(40, 33){\text{(d)}}

}
\end{picture}
\caption{(a) Evolution of the instantaneous energy of the bare system $\langle \mathcal{H}_0 \rangle$ during the adiabatic phase while slowly switching off the magnetic field. (b) The overlap between the instantaneous state and the ground state of the bare system, $|\langle \psi_0 | \psi(t) \rangle|^2$. Insets (c) and (d) show the value of $B(t)$ as it changes from 10 to 0 and the change of the total magnetization of the system with time as the field is slowly switched off respectively. The initial spikes in (a) and (d) correspond to the initial phase of polarizing the system along the external magnetic field.}

\end{figure}

\begin{figure}[t!] \setlength{\unitlength}{0.1cm}
\begin{picture}(60 , 65 )
{
\put(28, 0){ \includegraphics[scale=0.35]{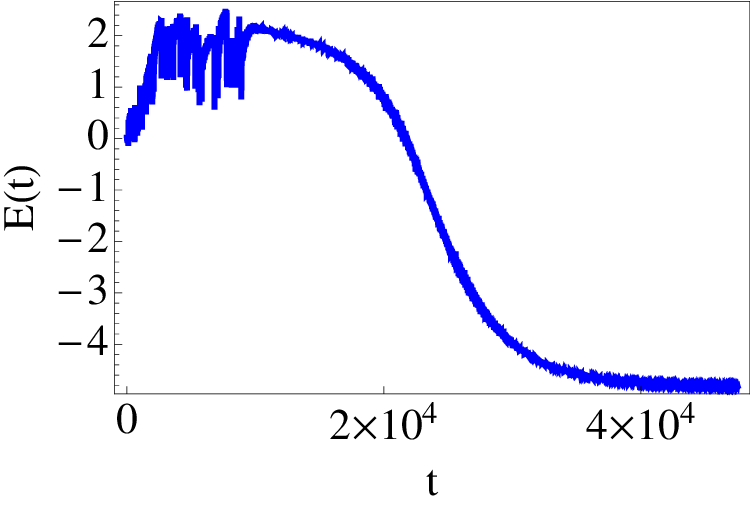}}
\put(-15, 0){\includegraphics[scale=0.33]{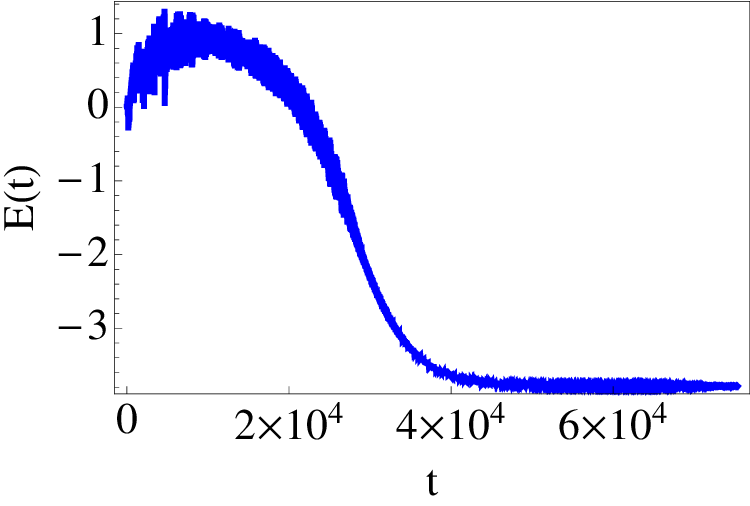}}

\put(28, 33){ \includegraphics[scale=0.35]{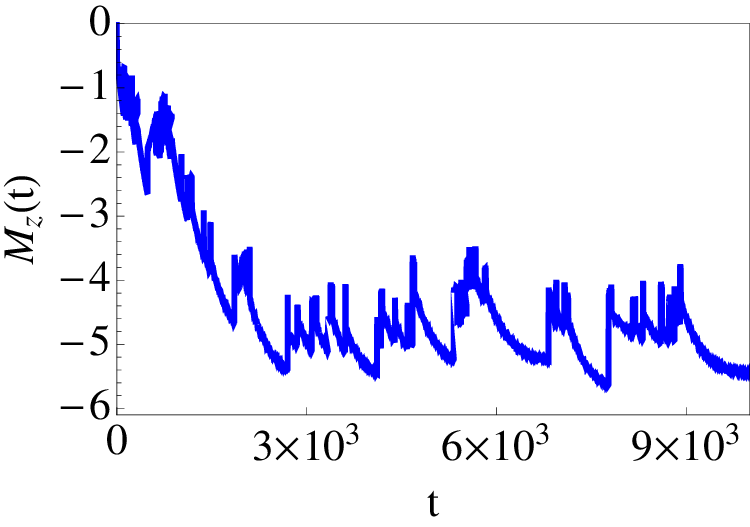}}
\put(-15, 33){\includegraphics[scale=0.34]{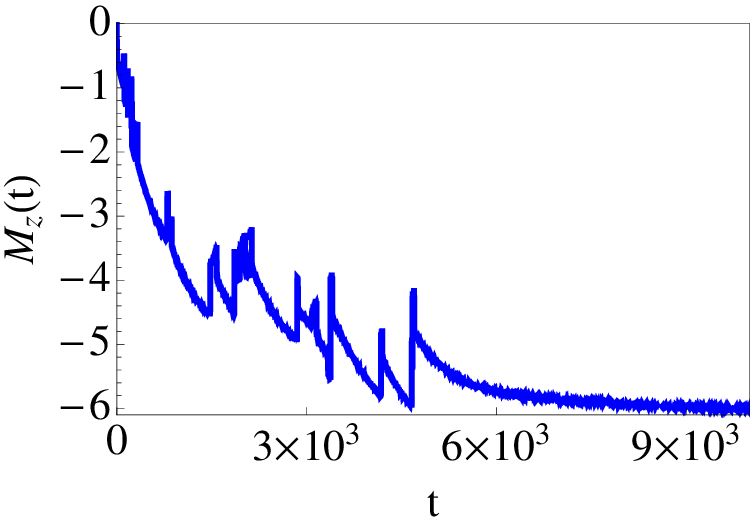}}

\put(22, 58){\text{(a)}}
\put(66, 59){\text{(b)}}

\put(21, 24.5){\text{(c)}}
\put(66, 26){\text{(d)}}

}
\end{picture}
\caption{(a,b) Evolution of the total magnetization for quantum spin chains consisting of 14 spin-1/2 particles with short-range and long-range interactions respectively while applying the scheme of repetitively measuring one single probe spin to polarize the system along an external magnetic field $\boldsymbol{B}$ (Step I). (c,d) Evolution of the instantaneous energy of the bare Hamiltonian $\langle \mathcal{H}_0 \rangle$ for both systems during the adiabatic evolution phase while slowly switching off $\boldsymbol{B}$. The adiabatic evolution process (Step II) starts at $t=10^4  \text{ s}$. }

\end{figure}
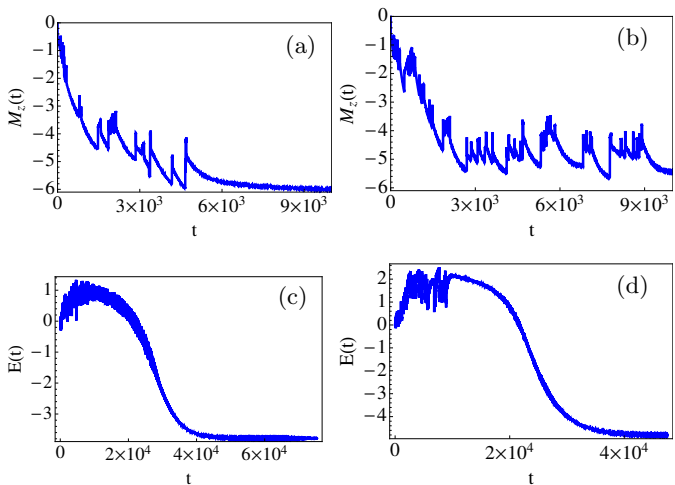

\subsubsection{Fixed single-particle probe}
The proposed algorithm may be difficult to implement when there is a practical constraint that makes it hard to selectively measure every single particle; for example, in an optical lattice where single-site measurement is performed by an off-resonant laser beam  that cannot be steered along the lattice with sufficient resolution \cite{gross2015}. Nevertheless, we can still use a variant of the proposed scheme while measuring  one single particle that will serve as a probe for the whole system.

 The core idea is to polarize that particular spin in the desired direction by the act of measurement and then let it unitarily interact with the rest of the system for a short time interval. During the unitary interaction, the probe spin will transfer part of its polarization to the rest of the system before we polarize it again by a new projective measurement and repeat this procedure many times. Upon measurement, given that the time interval of the unitary evolution is short enough, the probe spin will be projected to the desired direction with a high probability. If it is projected in the undesired direction, parallel to $\boldsymbol{B}$, we keep perturbing the system with RF pulses interspersed by acts of projective measurements of that spin, as we did in the first scheme, till we project it along the desired direction. Note that the time interval between successive measurements should be shorter than the characteristic timescale of the intrinsic dynamics of the system governed by $\mathcal{H}_0$. Otherwise, it will have equal probabilities of being projected in either direction upon each measurement. On the other hand, as in the previous scheme, it shouldn't be much shorter than the intrinsic dynamics timescale in order to avoid the onset of the quantum Zeno effect, which would freeze out the measured spin. After a large number of these measurement-interaction cycles, the system will be largely polarized opposite to $\boldsymbol{B}$. In Step II, the magnetic field will be switched off slowly as in the first scheme.

Because we take a single spin to be representative of the entire system, it is expected that this scheme will work best when the system exhibits translation-invariance. During Step I,  the strong magnetic field plays an important role in making the total polarization a quasi-conserved quantity. This property assists in transferring the polarization from the probe spin to the rest of the system through the inter-particle interaction during the unitary evolution periods between the repetitive acts of measurements. Note that in actual NMR experiments of systems of interacting magnetic dipoles, a very strong field is typically used, and by working in the rotating reference frame defined by the Larmor frequency, the dipole-dipole interaction Hamiltonian reduces to the so-called truncated Hamiltonian which exactly conserves the total magnetization along the magnetic field direction \cite{abragam1961}.

Let us demonstrate the efficiency of this scheme by applying it to the same system above with short-range interactions and another system with long-range interactions. The Hamiltonian for the latter is described by: \[\mathcal{H}_0=\sum_{m<n} J_{mn}  \bigl(S_m^x S_n^x-0.5(S_m^y S_n^y-S_m^z S_n^z) \bigr)+0.3\sum_{m}S_m^y,\] where the interaction strength $J_{mn}$ falls off inversely with the distance between spin $m$ and spin $n$ as $\frac{1}{|m-n|}$. The ground state energy of  this system is $E_0=-6.59  \text{ J}$. We show in Figs. 3-a and 3-b the evolution of the total polarization $M_z$ of both systems under the  sequence of measurements of one single spin and the occasional RF pulses. The spikes in $M_z$ correspond to instances where the probe spin is occasionally projected in the undesired direction, while the smooth sections in the plots correspond to the time intervals of pure repetitive measurements without applying the RF excitation pulses. During these smooth intervals, the sequence of measurements and unitary evolution steadily increases the total polarization of the system. The drawback of this scheme is the  considerably longer time taken to polarize the system compared with the first algorithm that measures all the spins.

We notice in Fig. 3 that, for the first Hamiltonian with short-range interactions, we could achieve a polarization strength of 85\% of the maximum polarization while, for the second Hamiltonian with long-range interactions, we achieved 70\% of the maximum polarization.  After gradually switching off the magnetic field starting at $t=10^4  \text{ s}$ with a decay constant $T_0=8\times 10^3  \text{ s}$  for both systems, we observe in Figs. 3-c and 3-d that they settle at energies \mbox{-3.78 J} and \mbox{-4.81 J},  which differ from $E_0$ by  10\% and 27\%, respectively. The fidelity of the final state with respect to the ground state in both cases is of the order of 10\%.

\subsection{Coarse measurement by a bulk magnetization detector}
Measuring a single particle may not be feasible in many situations. A more realistic approach involves using a probe that measures the collective magnetization from several spins, which constitute a subset of the whole system. In this part, we analyze the implementation of the algorithm introduced earlier using a coarse-grained probe that can only measure the direction of the magnetic field produced by $\boldsymbol{n}$  spins. Here, the probe is only sensitive to the direction of the magnetization and produces binary results depending on whether the magnetization of the measured subset of the system is positive or negative (along or opposite to the direction of $\boldsymbol{B}$). As before, these binary results are then used to switch on or off the RF field. This measurement essentially probes a single degree of freedom. 
The state of system is assumed to be minimally perturbed by this measurement and the wavefunction $|\psi\rangle$ is projected onto the subspace corresponding to the measured value (i.e., if  the measured value is positive, the state is projected onto the subspace having the majority  of the $\boldsymbol{n}$ spins measured by the probe oriented along $\boldsymbol{B}$ and vice versa). This subspace has a size equal to $2^{N-1}$.
As with the single-particle probe, the bulk detector can be positioned at a fixed location or configured to randomly jump throughout the system. 


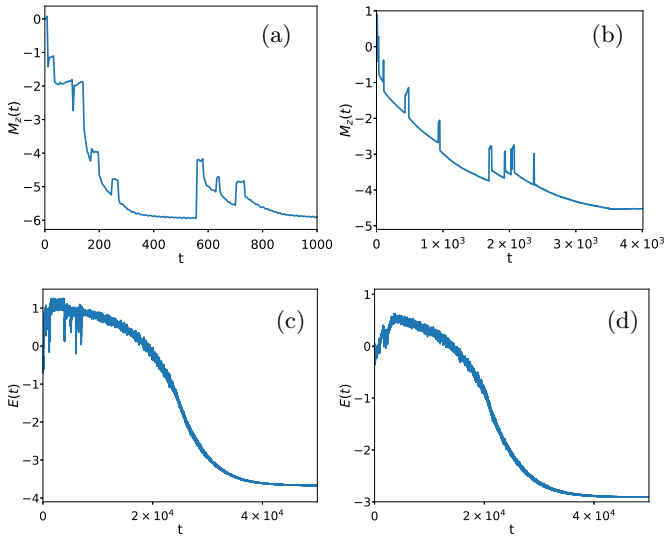
\begin{figure}[t!] \setlength{\unitlength}{0.1cm}
\begin{picture}(60 , 70 )
{

\put(-15,0){\includegraphics[scale=0.1]{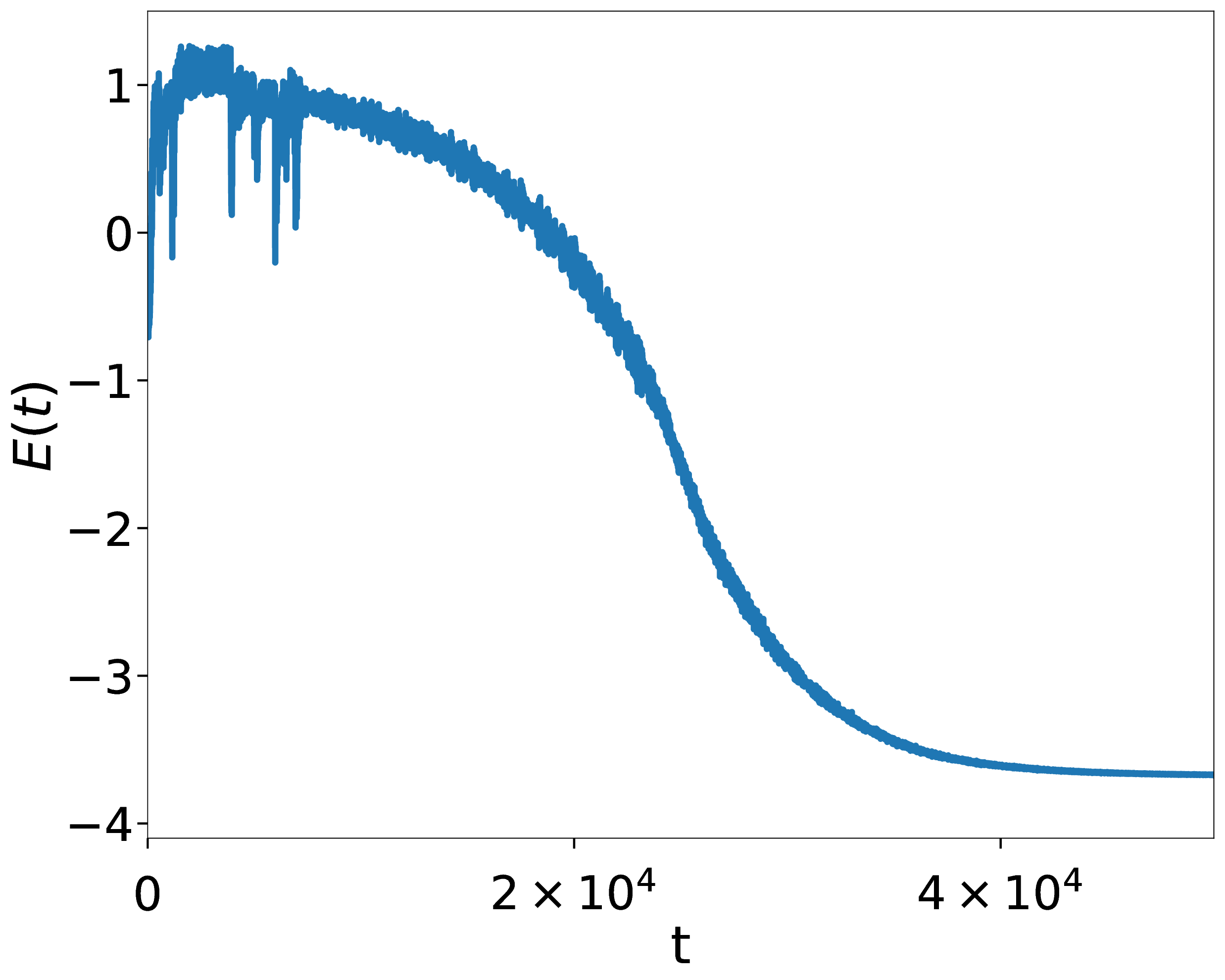}}
\put(28, 0){ \includegraphics[scale=0.1]{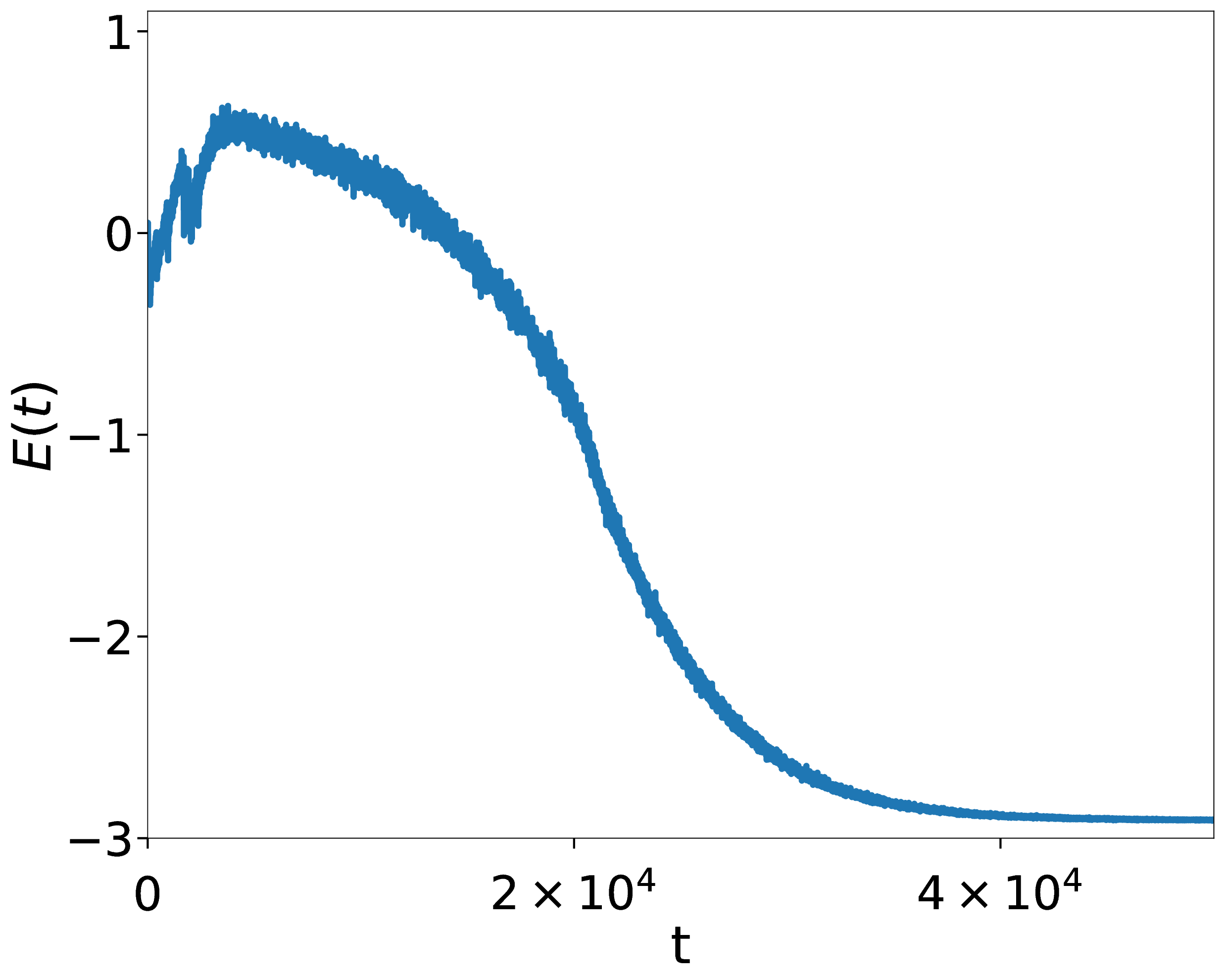}}

\put(-15, 36){\includegraphics[scale=0.105]{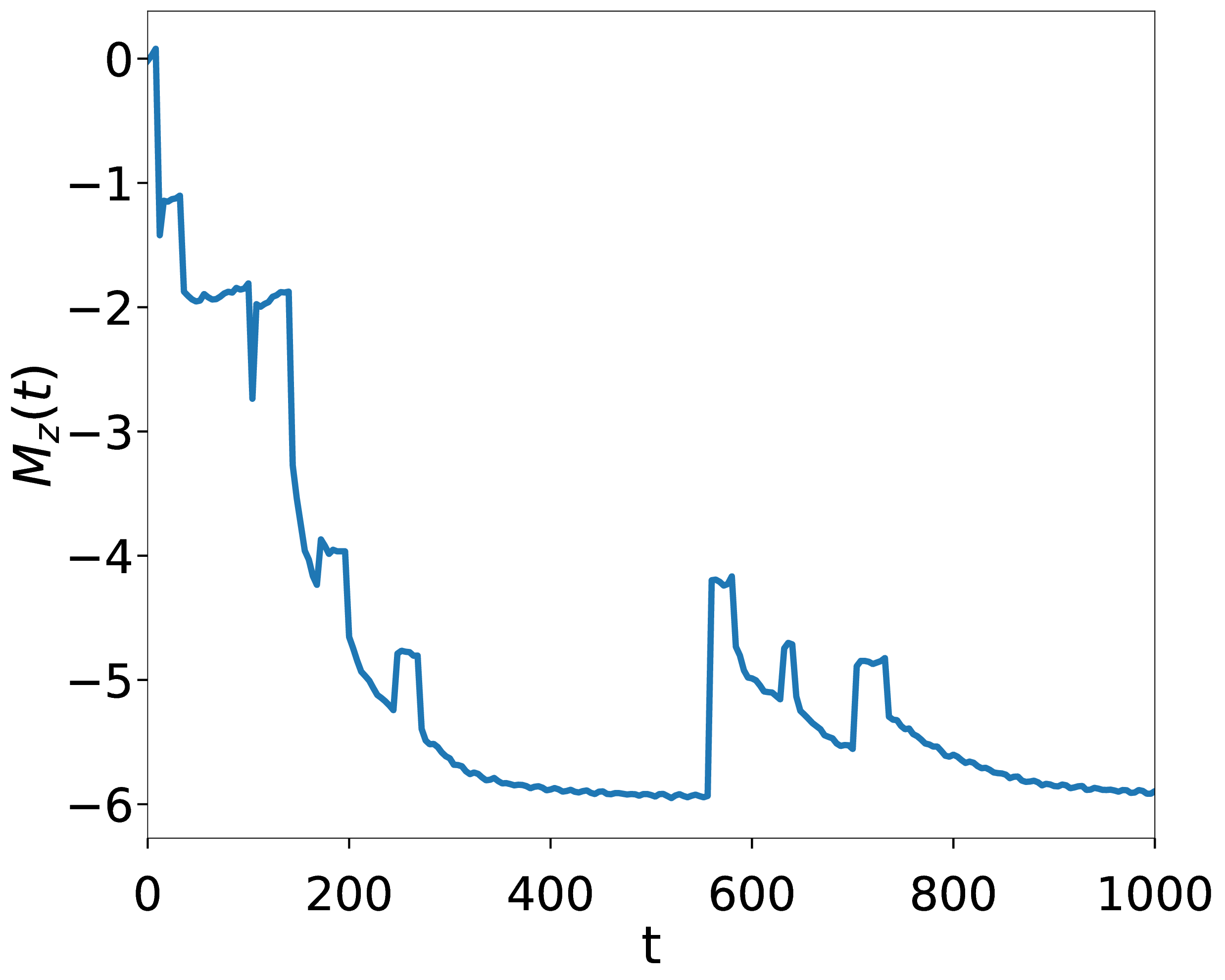}}
\put(28, 36){ \includegraphics[scale=0.105]{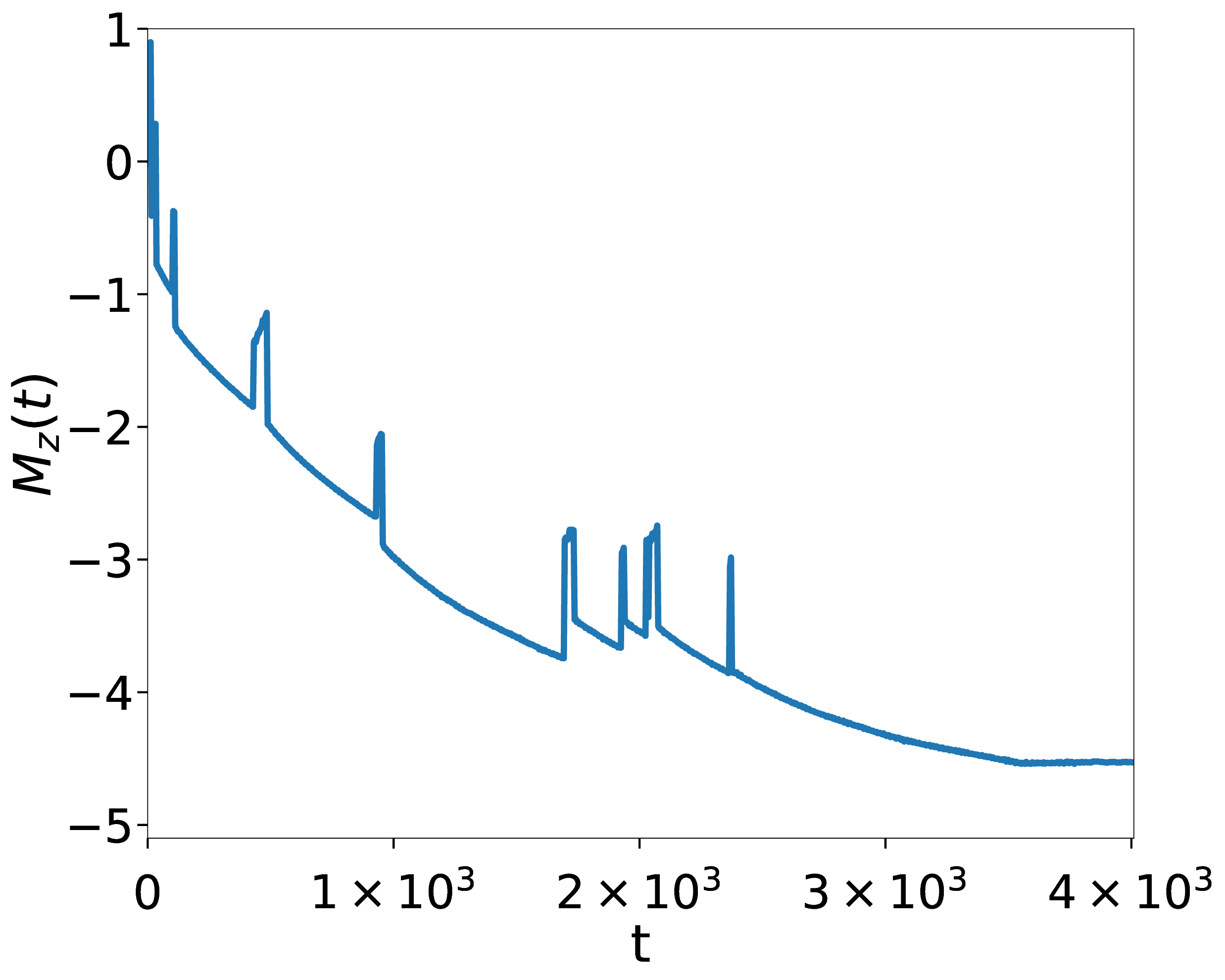}}

\put(19, 66){\text{(a)}}
\put(63, 66){\text{(b)}}

\put(21, 28){\text{(c)}}
\put(65, 28){\text{(d)}}

}
\end{picture}
\caption{(a,b) Evolution of the total magnetization of a system consisting of 14 spin-1/2 with short-range interactions during the polarizing phase of the algorithm (Step I),  using bulk probes consisting of 3-spin subsets.  In (a), the probe is randomly jumping through the chain, while in (b) the probe is fixed at a specific location.  
(c,d) Evolution of the instantaneous energy of the bare Hamiltonian, $\langle \mathcal{H}_0 \rangle$ for the two systems respectively during the adiabatic evolution of the system (Step II) while the external field $\boldsymbol{B}$ is being slowly switched off. }

\label{}
\end{figure}

We present in Fig. 4 the results of the algorithm applied on the same system from the previous section having short-range interactions, but now using a coarse-grained probe with $\boldsymbol{n}=3$. The probe randomly jumps throughout the system in Fig. 4-a, 4-c and is fixed at a certain location in Fig. 4-b, 4-d. A time step $dt=0.002 \text{ s}$ is used in this section. To illustrate the effect on the wavefunction due to a single act of measurement, consider a  scenario where the probe measures the first three spins in the chains.   In this case,  $|\psi\rangle$ is either projected onto the subspace
 $\{ | \uparrow \uparrow \downarrow \cdots  \rangle,
 |\uparrow \downarrow \uparrow \cdots  \rangle,
| \downarrow \uparrow \uparrow \cdots  \rangle,
| \uparrow \uparrow \uparrow \cdots  \rangle \}$ or 
 $\{ | \downarrow \downarrow \uparrow \cdots  \rangle,
 |\downarrow \uparrow \downarrow \cdots  \rangle,
| \uparrow \downarrow \downarrow \cdots  \rangle,
| \downarrow \downarrow \downarrow \cdots  \rangle \}$,  depending on the measurement outcome. 
We notice that the maximum polarization obtained by the bulk probe of 3 spins is worse than the single-particle probe of Section A, and thus the lowest energy achieved is further from the true ground state energy. The efficiency gets worse as the size of the probe ($\boldsymbol{n}$) increases and should not, in general, exceed half the system size.  In other words, there is a trade-off between the measurement probe granularity and the achievable fidelity, where larger probe sizes lead to less efficient cooling.

 In Fig. 5, we show the polarization achieved by using bulk probes that measure subsets of 6 spins in spin chains consisting of 18 spins (i.e., the probe measures a third of the whole system at a time). Here,  configurations corresponding to zero magnetization are also considered to be favorable states alongside the anti-parallel polarization states. The results demonstrate that with a randomly jumping probe (as shown in Figs 5-a and 5-c), 75\% of the maximum polarization is achieved. In contrast, a fixed probe (Figures 5b and 5d) reaches only 50\% of the maximum polarization. We hope that this method will open the door to employing more coarse probes that can measure collective degrees of freedom of macroscopic size spanning the entire system such as the total magnetization of the system  (see also the discussion in \cite{elsayed2022}).


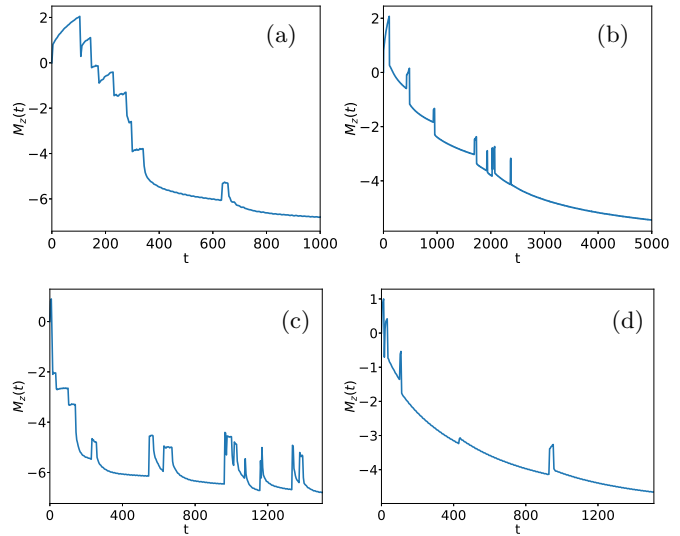
\begin{figure}[t!] \setlength{\unitlength}{0.1cm}
\begin{picture}(60 , 70 )
{

\put(-15,0){\includegraphics[scale=0.1]{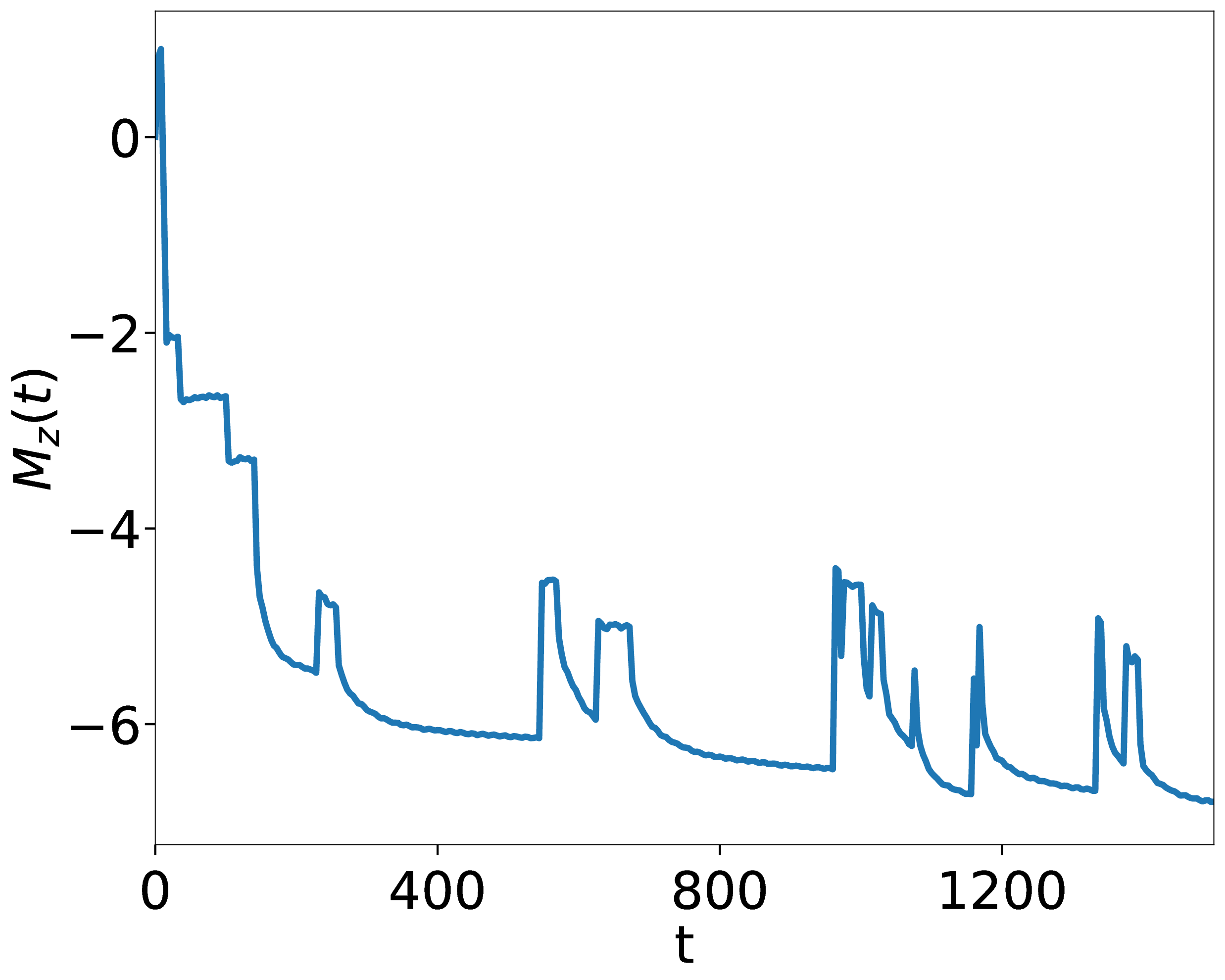}}
\put(28, 0){ \includegraphics[scale=0.1]{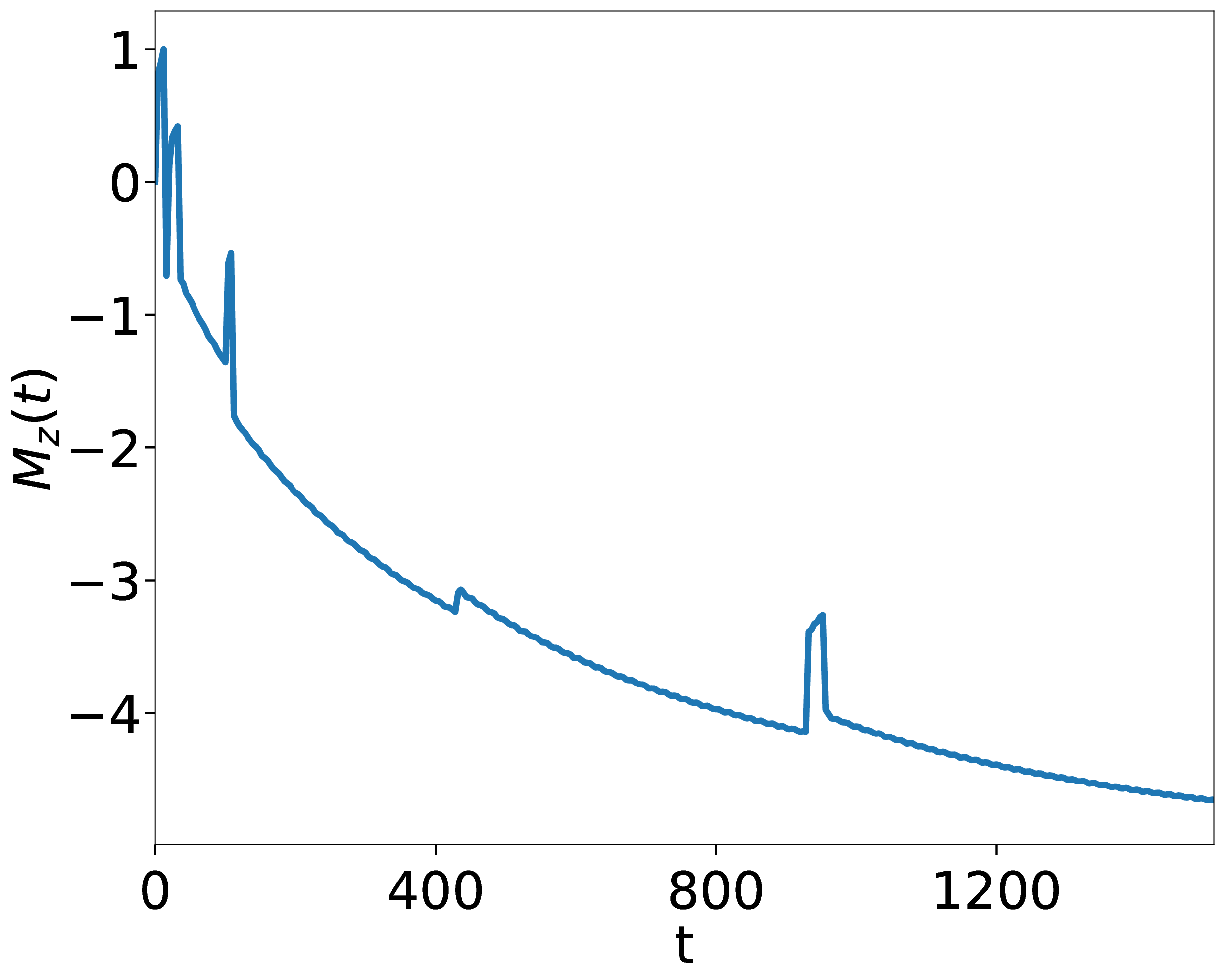}}

\put(-15, 36){\includegraphics[scale=0.105]{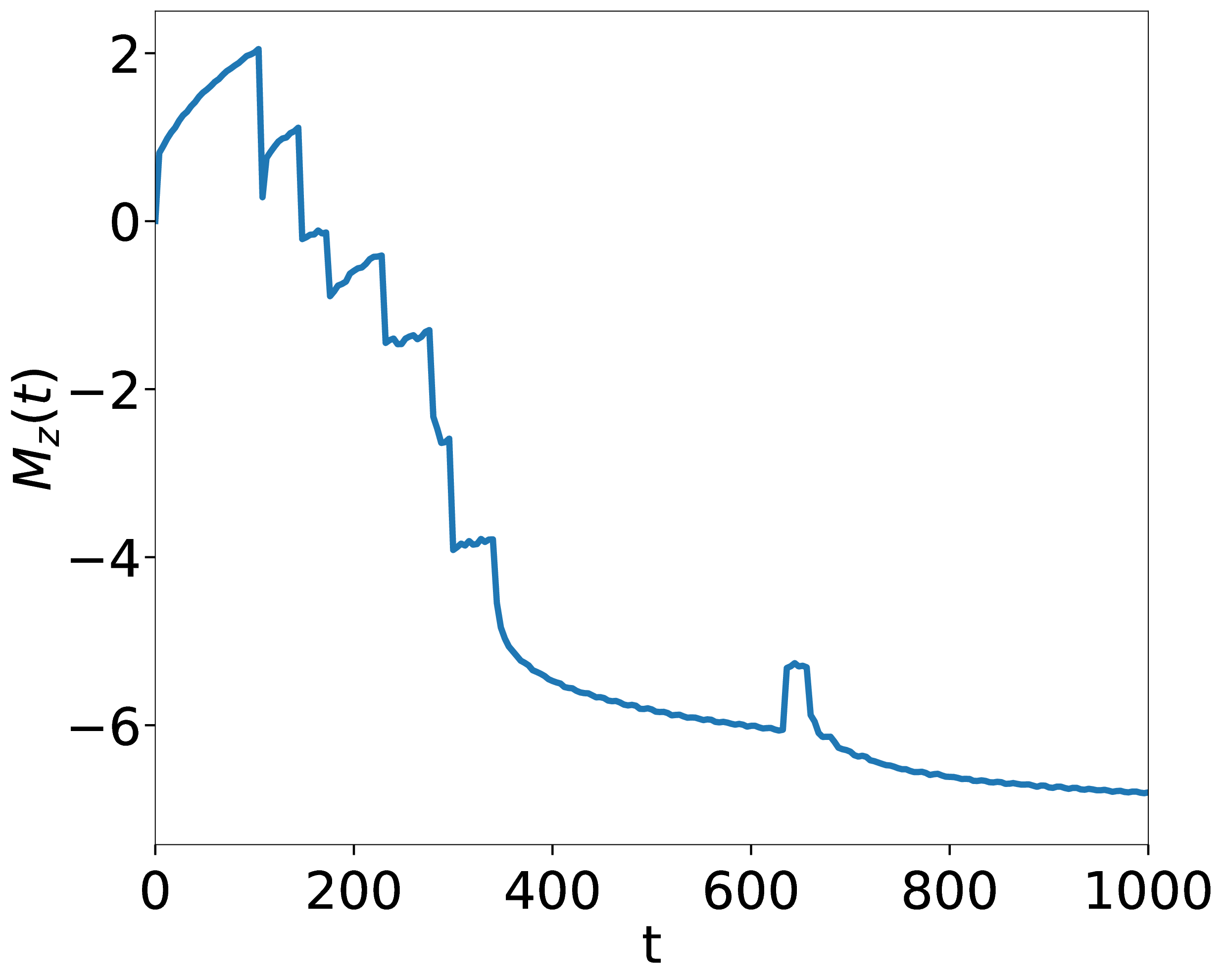}}
\put(28, 36){ \includegraphics[scale=0.105]{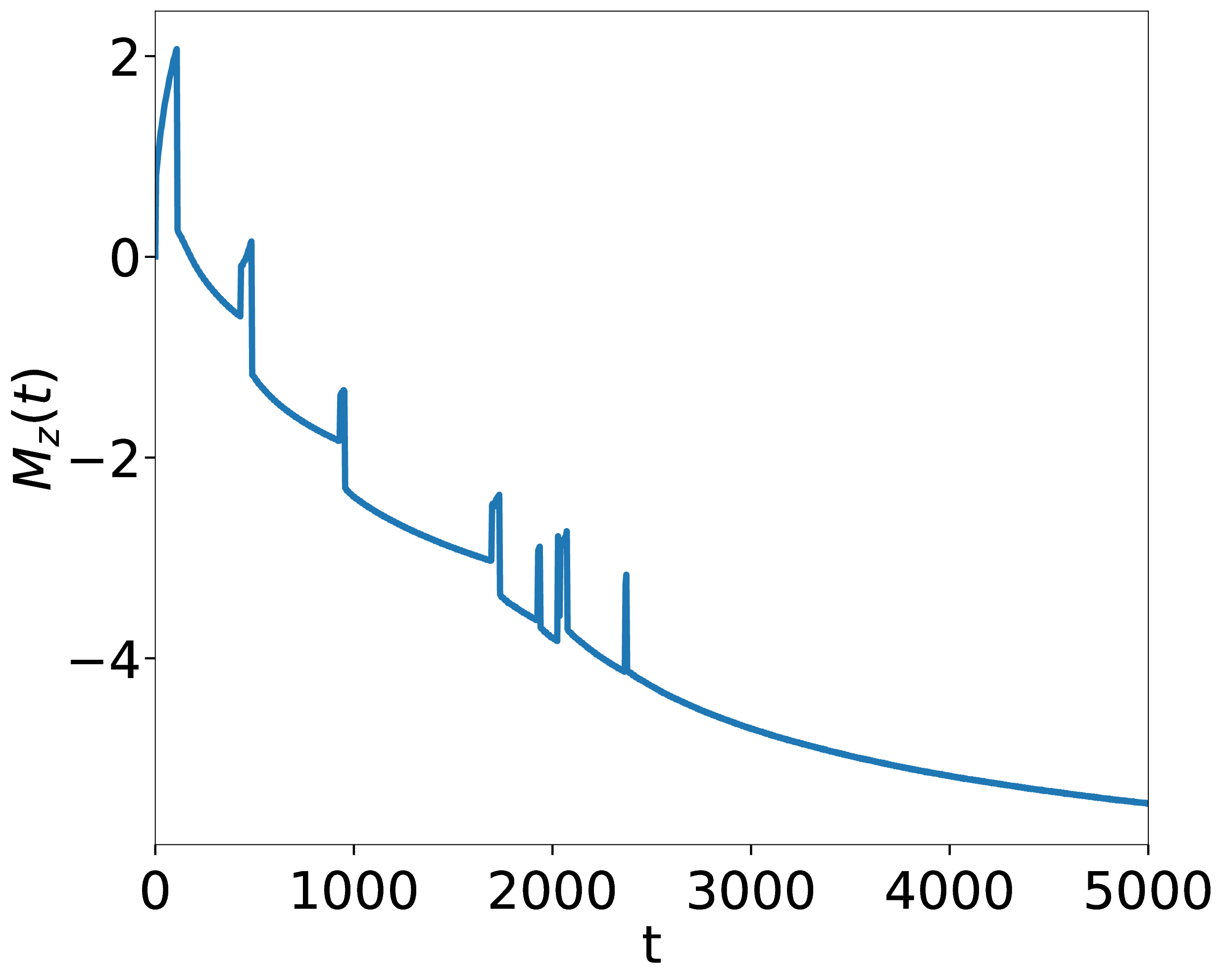}}

\put(19, 66){\text{(a)}}
\put(63, 66){\text{(b)}}

\put(21, 28){\text{(c)}}
\put(65, 28){\text{(d)}}

}
\end{picture}
\caption{Evolution of the total magnetization of a system consisting of 18 spin-1/2 while applying the measurement-based polarization algorithm starting from an unpolarized state and using bulk probes of 6-spin subsets  measuring the direction of the magnetization (Step I). (a,b) The system involves short-range interactions and the measurement is done using a jumping probe (a) and a fixed probe (b). (c,d) The system involves long-range interactions and the measurement is done using a jumping probe (c) and a fixed probe (d).  }

\label{}
\end{figure}

\section{Concluding remarks}
To conclude, we have presented a cooling technique that effectively cools down a many-body quantum system described by an unknown Hamiltonian in an arbitrary state to a state close to its ground state using a combination of measurement-based polarization and adiabatic demagnetization. The basic principle is a divide-and-conquer strategy, where we avoid the difficult task of polarizing the entire system directly -- which would require exploring the full Hilbert space of the system -- and replace it by the easier task of polarizing small segments of the system, one at a time, which requires exploring a much smaller portion of the Hilbert space. The latter task is easily achieved by the mere act of successive measurements using either single-particle probes or multi-particle bulk probes, which are sensitive only to the direction of the magnetization. The first version of the technique requires measuring every particle (or subset of particles) of the system till it is guaranteed that the majority of the particles are polarized anti-parallel to a strong external field. The second version measures one single particle (or a subset of particles) successively, allowing it to transfer its polarization to the rest of the system during the unitary evolution between the successive acts of  measurements. In both cases, RF pulses are employed to perturb the system if the measurement outcome is along the undesired direction till it is projected onto the desired direction. Switching off the field slowly afterwards allows the system to adiabatically approach a state very close to its ground state.

The fidelity of the final state depends on three factors: (i) The polarization achieved at the end of Step I, which depends on the measurement technique used and the parameters of the control scheme. (ii) The closeness of the polarized state to the true ground state of the polarized system, which in turn depends on the strength of the external magnetic field with respect to the typical interaction strength of the intrinsic Hamiltonian of the system. (iii) The accuracy of the adiabatic evolution phase. It is expected that the adiabatic evolution phase will suffer from several practical constraints, which can impact its effectiveness  \cite{bukov2018}. Depending on the phase structure of the system, the timescale required for achieving a truly adiabatic driving may diverge and therefore,  make it challenging to obtain a very high fidelity if the system exhibits a quantum phase transition or the energy levels exhibit avoided crossings  \cite{jensen2021,doria2011}. These factors will limit the maximum fidelity that can be achieved in real situations or, alternatively, limit the type of systems for which this technique can be used.

The half-cycle measurement frequency used in our technique poses a practical challenge on the required RF switching frequency. While recent advances in ultrafast RF switching techniques \cite{ge2015} may mitigate this problem, we leave it for further research to investigate a broader parameter space for our technique to work optimally while having less stringent requirements on the measurement and excitation apparatus.

In case simultaneous measurements of many particles are feasible, an effective strategy to speed up the polarization step (Step I) is to employ a repetitive measurement of each spin once it is polarized along  the desired direction, as an add-on to the original algorithm, to confine that spin along that direction using quantum Zeno dynamics as in \cite{sorensen2018}. Another variation of Step I involves adjusting the strength of the RF field based on  the proximity to the desired state, i.e., to reduce the strength of the perturbation the higher the number of consecutive measurements outcomes are obtained in the desired direction along the external field. This approach aims to reduce the probability of flipping a spin that has already been confirmed to align in the desired direction.

The numerical simulation  of interacting quantum spin systems presented in this work serves as a proof-of-concept for the validity of the algorithm; however, the technique is generic and can be extended to other strongly correlated systems such as ultracold atoms in an optical lattice. These systems are effectively isolated from the environment and thus the unitary dynamics underlying our technique is still relevant. Moreover, all the components of the proposed technique have already been incorporated in the control of ultracold atom systems. For example, the back-action induced by the act of measurement has  been shown to be an effective tool for steering the system towards desired unconventional states \cite{patil2015,langbehn2023}. Additionally, it has  been demonstrated that adiabatic demagnetization can be an effective cooling technique of ultracold atoms \cite{medley2011,schachenmayer2015,mirasola2018}. Recent advances in the ability to address individual atoms in an optical lattice by advanced imaging and detection techniques such as the quantum microscope \cite{ott2016, gross2021} make the proposed technique of local measurements a viable approach in these systems. While, in the context of NMR experiments, RF pulses are used  to `shake up' the system after the measurements, other means of perturbation can be employed in other contexts. For instance, in ultracold atoms in optical lattices, we can vary the parameters of the lattice or the interatomic interactions or use artificial gauge fields \cite{schafer2020}. 

It remains an open question whether the path from the infinite temperature initial state to the polarized state in Step I, going through a sequence of measurements and perturbations, corresponds to imaginary time evolution governed by $\mathcal{H}$. If this were the case, the partially polarized states achieved along this path would correspond to states sampling the Gibbs distribution $\rho\approx e^{-\beta \mathcal{H}}$, where $\beta$ is the inverse temperature. This correspondence would suggest potential in using the  proposed technique in the preparation of thermal states for quantum simulators as in \cite{motta2020}.
 Quantum simulation is concerned with using quantum computers to simulate atomic and molecular systems. Studying the thermal properties of materials using a quantum computer often requires the preparation of Gibbs states which is usually done using the phase estimation algorithm \cite{poulin2009}. However, this algorithm typically requires very large quantum circuits \cite{powers2023}. 
Our measurement-based algorithm can offer a promising alternative. In order to generate finite-temperature states, we can allow the state of the quantum circuit at the end of Step I to be a partially polarized state. Starting the adiabatic evolution in Step II by such a state with finite entropy would  lead to a finite-temperature state at the end of Step II. In a quantum computer setting, the degree of controllability over measuring or manipulating individual qubits is far better than in conventional physical systems, and hence we anticipate that the proposed technique will exhibit much better performance when implemented in quantum simulators.

\section{ACKNOWLEDGMENT}
The author thanks Intel Corporation for granting him access to the  Intel\textsuperscript{\tiny\textregistered} DevCloud facility where the numerical simulations presented in this article were performed  as one of the oneAPI projects.

\bibliographystyle{apsrev4-1}
%
\end{document}